\documentclass[12pt]{article}
\usepackage{graphicx,amsmath,amssymb,cite}
\usepackage{epsfig,epsf}
\usepackage{graphicx}
\usepackage{epstopdf}
\newcommand{\beq}{\begin{equation}}
\newcommand{\eeq}{\end{equation}}
\newcommand{\alphabar}{\bar{\alpha_s}}
\newcommand{\betabar}{\bar{\beta_0}}

\numberwithin{equation}{section}

\begin{document}

\vspace*{0.5 cm}

\begin{center}

{\Large{\bf The Green Function for the BFKL Pomeron and the Transition to DGLAP  Evolution. }}

\vspace*{1 cm}

{\large H. Kowalski~$^1$, L.N. Lipatov~$^{2}$, and D.A. Ross~$^3$} \\ [0.5cm]
{\it $^1$ Deutsches Elektronen-Synchrotron DESY, D-22607 Hamburg, Germany}\\[0.1cm]
{\it $^2$ Petersburg Nuclear Physics Institute, Gatchina 188300, St. Petersburg, Russia}\\[0.1cm]
{\it $^3$ School of Physics and Astronomy, University of Southampton,\\Highfield, Southampton SO17 1BJ, UK}\\[0.1cm]
 \end{center}

\vspace*{3 cm}

\begin{center}
{\bf Abstract} \end{center}
 We consider the (process-independent) Green function for the BFKL equation
 in the next-to-leading order approximation,
 with running coupling, and 
explain how, within the semi-classical approximation, it is related to Green function of the Airy equation.
The unique Green function is obtained from a combination of its required ultraviolet
behaviour compatible with asymptotic freedom and  an infrared limit phase imposed by the non-perturbative
sector of QCD. We show that at sufficiently large gluon transverse momenta the corresponding
gluon density matches that of the DGLAP analysis,
whereas for  relatively small
 values of the gluon transverse momentum the gluon distribution is sensitive to the  Regge poles,
 whose positions are determined both by the non-pertubative QCD dynamics and  physics
at large transverse momenta.


\vspace*{3 cm}

\begin{flushleft}
  January 2014 \\
\end{flushleft}

\newpage

\section{Introduction}

In recent papers \cite{EKR,KLRW,KLR}, it has been shown that the discrete BFKL
\cite{BFKL} pomeron can reproduce the low-$x$ structure functions at HERA very well,
by properly determining the oscillation phases of wavefunctions  at the infrared boundary.
The discrete pomeron arises from the accounting for the running of the coupling
with gluon transverse momentum and the imposition of such phases (see 
 \cite{lipatov86}).
 It was furthermore shown
  that the quality of the fit is sensitive to the exact
 $\omega$-plane positions of third and higher Regge poles which are influenced (according to the BFKL equation)
by hypothetical heavy particles and their interactions 
\cite{KLR}. Note that the thresholds for such particles beyond the Standard Model (BSM) may be above the energy
scale at which the structure functions are measured. Such sensitivity of the pomeron spectrum is similar to the sensitivity of
 the  weak-mixing angle, $\theta_W$, to different  Grand Unified Theories (GUTs). The pomeron spectrum, $\omega_n$,
and the corresponding complete set BFKL eigenfunctions, satisfying appropriate boundary conditions,
 $f_n(t)$, determine the Green function
  \beq {\cal G}(t,t^\prime,Y)  \   = \ \sum_{n=1}^\infty e^{-\omega_n Y} f_n(t) f^*_n(t^\prime) . \eeq
The main problem with this representation is a very slow convergence of the sum over pomeron contributions, so that in refs.\cite{KLRW,KLR} it was necessary to take a very large number ($> \, 100$)  of them
in order to obtain a good desciprtion of the data. One of the purposes of this paper is to find an alternative
representation of the Green-function which does not suffer from this disadvantage.

The use of the Green function approach enables the calculation of the ampltidue for each specific process
(such as structure fiunctions in deep-inelastic scattering)
solely as a  convolution
of this Green function with  impact factors that encode the coupling of the
 Green function to the external particles that participate in that process. Thus,
 for example, the structure function, $F_2(x,Q^2)$ at low-$x$ is given by
 \beq F_2(x,Q^2)  \ = \ \int dt dt^\prime \Phi_{DIS}(Q^2,t) {\cal G}(t,t^\prime,Y)
  \Phi_P(t^\prime), \eeq
where, $Y=\ln(1/x)$, $t=\ln(k^2/\Lambda_{QCD}^2), \ t^\prime=\ln(k^{\prime \,2}/\Lambda_{QCD}^2)$; 
$k, \ k^\prime$ being the transverse momenta of the gluons entering the
 BFKL  amplitude.
 $\Phi_{DIS}(Q^2,t)$ describes the (perturbatively calculable) coupling of the gluon with transverse momentum $k$
 to a photon of virtuality $Q^2$ and $\Phi_P(t^\prime)$ describes
the coupling of a gluon of transverse momentum $k^\prime$ to the target proton.

In the discrete version of the BFKL formalism the Mellin transform of the 
Green function
 $$ {\cal G}_\omega(t,t^\prime) \  \equiv \ \int dY e^{-\omega Y}
 {\cal G}(t,t^\prime,Y), $$
for positive $\omega$, has a set of poles  at $\omega=\omega_n$
(as opposed to a cut along the real axis in the case where there is no 
restriction on the infrared behaviour of the BFKL amplitude).

We can define the Mellin transform of the unintegrated gluon density,
 ${\cal A}_\omega(t)$ as the convolution of the Mellin transform of the Green
function with the proton impact factor
\beq {\cal A}_\omega(t) \ \equiv  \ \int dt^\prime {\cal G}_\omega(t,t^\prime)
 \Phi_p(t^\prime) \label{amp} \eeq

This immediately poses the question as to how the results from the discrete BFKL
formalism can match those of a DGLAP analysis \cite{DGLAP}
 in 
DLL limit where both $Y$ and $t$ are large, but obey the inequality
 $$ \alpha_s(t)  Y \ \ll \ 1, $$ 
 for which the function
${\cal A}_\omega(t)$ obeys the DGLAP equation
 \beq e^{-t/2}  \frac{\partial}{\partial t}\left\{ e^{t/2} {\cal A}_\omega(t)
\right\}  \ = \ 
    \frac{C_A \alpha_s(t)}{\pi \omega}  {\cal A}_\omega(t) \label{DGLAP} . \eeq

In the case of the purely perturbative BFKL formalism with a cut
 singularity in $\omega$, this match is understood \cite{CC,abf}
 from the fact that at large $t$ and small $\omega$, the Mellin transform
function from the BFKL analysis is approximated by
 \beq {\cal A}_\omega(t) \   \sim \ \exp\left\{ 
 -\int^t \frac{C_A \alpha_s(t^\prime)}{\pi\omega} dt^\prime
\right\}, \eeq
which is a solution to eq.(\ref{DGLAP}) and the
unintegrated gluon density (i.e.  inverse Mellin transform
  of ${\cal A}_\omega(t)$) is dominated by a saddle-point at
 \beq \omega \ = \ \sqrt{\frac{C_A \alpha_s(t)}{\pi Y}}. \label{saddle99} \eeq

In this paper, we show that provided the Green function is carefully defined
and its boundary conditions adequately specified, then at sufficiently
large gluon virtuality, a similar matching occurs. In section 2, we discuss the
semi-classical approximation for the Green function of the BFKL equation, 
without reference to any specific process and in section 3 we consider its 
application to deep-inelastic scattering and discuss under what circumstances
 we expect a match to the result of a DGLAP analysis in the 
double-leading-logarithm (DLL) limit.

\section{The BFKL Green Function}
Our approach to the BFKL equation is similar to the DGLAP approach with the difference that instead of the first order differential equation in $t$, as we have  in the DGLAP case,  we will 
write a simplified BFKL equation as
  a second order differential equation, which could be considered as a  quantized version of  the DGLAP equation.

In general, the BFKL Green function (in Mellin space) ${\cal G}_\omega(t,t^\prime)$
(with appropriate boundary conditions)
  obeys the equation
\beq \left( \omega- \hat{\Omega}(t,\hat{\nu}) \right) {\cal G}_\omega(t,t^\prime)
 \ = \ \delta(t-t^\prime), \label{BFKLeq} \eeq
where $\hat{\Omega}$ is the (Hermitian) BFKL operator (with running coupling)
and
 $\hat{\nu} \, \equiv -i \partial/\partial t$ is the operator conjugate
to $t$. In the LO approximation
(and neglecting quark masses) the BFKL  operator  is given in terms of the 
leading order expression for the
characteristic function, $\chi(\alpha_s(t),\nu)$, by
\beq \hat{\Omega} \ = \ \frac{1}{\sqrt{\betabar t}} \left( 
 2\Psi(1) - \Psi\left(\frac{1}{2}+\frac{\partial}{\partial t} \right)
-\Psi\left(\frac{1}{2}-\frac{\partial}{\partial t} \right) 
 \right)  \frac{1}{\sqrt{\betabar t}}, \label{eqomop}, \eeq
 where we have used the notation  
$ \alphabar \, \equiv \, C_A \alpha_s/\pi$,  $\betabar \, = \,  \beta_0 \pi/C_A.$  
The hermiticity of the operator is assured by placing $\sqrt{\alphabar(t)}$ on either side
of the hermitian  differential operator.

Beyond leading order, the characteristic function, $\chi$, acquires an explicit $\omega$ dependence 
due to the summation of collinear divergences \cite{salam}. In this case, the quantity
 $\Omega(t,\nu)$ is obtained
from the solution to
 \beq \Omega \ = \ \chi\left(\Omega,\alphabar(t),\nu\right). \eeq
If $\Omega$ is expanded as a power series in $\alpha_s$ the result up to
order  $\alpha^2_s$  coincides with the NLO characteristic function \cite{FL}.
The operator $\hat{\Omega}$ is constructed by promoting the variable $\nu$ to the operator
$\hat{\nu}$ defined above, and symmetrizing as necessary in order to
generate a Hermitian operator \footnote{There is some ambiguity in the ordering
of operators for the construction of a Hermitian operator, but this ambiguity
does not affect the solution of the eigenvalue problem in the semi-classical
approximation.}.  
 The function, $\Omega(t,\nu)$  must
be determined for all values of $t$ and $\nu$ by numerical methods. Importantly, however, we note that it is symmetric under $\nu \, \leftrightarrow \, -\nu$  - i.e. it depends on  $\nu^2$ - so that the 
operator $\hat{\Omega}$ contains only even derivatives with respect to $t$. 

For a given eigenvalue, $\omega$, we define the classical frequency, $\nu_\omega(t)$  by
 \beq  \omega \ = \ \chi\left(\omega,\alphabar(t),\nu_\omega(t)\right),  \label{classical} \eeq
 (the subscript $\omega$ serves as a reminder that this classical frequency is $\omega$-dependent
as well as $t$-dependent). For any positive value of $\omega$, there exists a critical value, $t_c$
of $t$ such that
 \beq \chi(\omega, \alphabar(t_c),0) \ = \ \omega \eeq
($t_c$ is also $\omega$-dependent). For $t \, < \, t_c$, the classical frequency,
$\nu_\omega(t)$ is real and the eigenfunctions of the operator $\hat{\Omega}$
are oscillatory functions of $t$, whereas for $t> t_c$ the classical frequency
is purely imaginary and the (physically acceptable)
 eigenfunction is a monotically decreasing function of $t$. Thus $t=t_c$ represents 
a turning point in the eigenfunctions of $\hat{\Omega}$.

In the neighbourhood of the turning point,  the BFKL equation (\ref{BFKLeq})
is known to  reduce to the Airy equation. To see this,  we first define  two related variables, 
$s_\omega (t)$ and $z(t)$  (both variables are implicitly dependent on $\omega$)  . The variable $s_\omega (t)$ denotes the corresponding classical  action and is defined as   
\beq
s_\omega (t) \ = \  \int _{t}^{t_c} dt'\,\nu _\omega (t')\, ,  \label{action}
\eeq
and the (real) variable $z(t)$,  defined as 
\beq
z (t)=   -  \left( \frac{3}{2} s_\omega (t) \right)^{\frac{2}{3}}   \label{zvar}
\eeq
which obeys the differential equation
\beq \frac{dz(t)}{dt} \ = \  \frac{\nu_\omega(t)}{\sqrt{-z(t)}} \eeq
with boundary value $z(t_c)=0$.
 Near the critical value of $t$, ($t \, \sim \, t_c$), we have
 \beq z \ \approx   \left(t-t_c\right) \,
 \left( \frac{2 \dot{\Omega}}{\Omega^{\prime\prime}}  \right)^{1/3}_{|t=t_c,\nu=0}
   \label{zapprox}, \eeq
where $\prime$ indicates partial differentiation with respect to 
$\nu$ and $\cdot$ indicates partial differentiation with respect to $t$.  To derive this relation we used the fact that near the critical point we can expand the BFKL function as, 
\beq \Omega\left(t,\nu_\omega(t)\right)  -  \omega \approx  \dot{\Omega} (t-t_c) + \Omega^{\prime\prime}\nu_\omega^2/2,  \label{tcapprox} \eeq
 corresponding to the diffusion approximation. 
By substituting $\nu\rightarrow \hat{\nu}$ and changing variables to $z$ (\ref{zapprox})
 the BFKL  operator, $\hat{\Omega}$, (for a given eigenvalue, $\omega$) is simply related  to the Airy operator
\beq \left(\omega - \hat{\Omega}\left(t, -i\frac{\partial}{\partial t} \right)\right)
\;  = 
 \; \left( \frac{\Omega^{\prime\prime}}{2} \right)^{1/3} \;   (\dot{\Omega})^{2/3} \; 
  \left(z - \frac{\partial^2} {\partial z^2}  \right)  \label{airyeq}. \eeq

\subsection{Generalized Airy Operator}
  We now show that, in  the semi-classical approximation, the BFKL operator
 can be  related  to  the ``generalized Airy operator'', 
 both in the vicinity of the turning point $and$ far away from it.  This means that in both cases 
we can generalize eq. (\ref{airyeq}) to
\beq \left(\omega - \hat{\Omega}\left(t, -i\frac{\partial}{\partial t} \right)\right)
 \ 
\approx \ \frac{1}{N_\omega(t)} \left(\dot{z} z 
- \frac{\partial}{\partial t} \frac{1}{\dot{z}} 
 \frac{\partial}{\partial t} \right)  \frac{1}{N_\omega(t)} \label{eqi}. \eeq
 The RHS of this equation denotes the generalized Airy operator.  Near $t = t_c$,  (where $\dot{z}$ becomes a  constant)
 the equation (\ref{eqi}) becomes exact, as can be seen
from eq.(\ref{airyeq}). For $t$ far away from $t_c$ we will derive eq.(\ref{eqi})  in the  semi-classical approximation in which it is
 assumed that $\nu_\omega(t)$ is 
a sufficiently slowly varying function of $t$ so  that 
\beq
\left|\frac{d}{dt} \ln \nu(t)\right| \ll 1,
\eeq
and we may neglect higher than the
first derivatives of $s_\omega (t)$ with respect to $t$.

 We begin by determining  the normalization function $N_\omega(t)$ in the region  $t \, \approx t_c$. Near $t_c$,
 where the approximations (\ref{zapprox}) and (\ref{tcapprox}) are taken as exact,
 eq.(\ref{eqi})   becomes exact with
\beq N_\omega(t) \ = \ \left|\frac{1}{2}\sqrt{\dot{\Omega}}\, \Omega^{\prime\prime}\right|^{-1/3}_{|t=t_c,\nu=0} \label{eqj1} \eeq
Using the fact that 
 $$  \Omega^{\prime\prime} \ \stackrel{\nu\to 0}{\rightarrow}  \frac{\Omega^{\prime}}{\nu}, $$
it will turn out to be convenient to re-express this as
\beq N_\omega(t)
 \ = \ \frac{(-z)^{1/4}}{\sqrt{\frac{1}{2}\left|\Omega^{\prime}\right|}}  \label{eqj} \eeq

In the next step we  consider the region where $t$ is far from $t_c$.
Noting that
\beq
 \left(\dot{z} z 
- \frac{\partial}{\partial t} \frac{1}{\dot{z}} 
\frac{\partial}{\partial t} \right)  \ = \ \dot{z} \left(z-\frac{\partial^2}{\partial z^2} \right),
 \label{eqd}  \eeq
 we see that the two eigenfunctions  of the operator $\hat{\Omega}$ with eigenvalue $\omega$ 
 are given, in the approximation of  eq.(\ref{eqi}), by   
 $$ N_\omega(t) A_i(z(t)) \ \mbox{and} \  N_\omega(t) B_i(z(t))  $$
where $A_i(z), \, B_i(z)$ are the two independent Airy functions. 

Thus, in order to relate   eqs. (\ref{airyeq}) and (\ref{eqi}) we seek a function $N_\omega(t)$ such that 
\beq \left(\omega - \hat{\Omega}\left(t, -i\frac{\partial}{\partial t} \right)\right)
 N_\omega(t) A_i(z(t)) \ = \ 0,  \label{eqc}\eeq
 valid in the semi-classical approximation, for values of $t$ far from $t_c$.

To  determine the function $N_\omega(t)$ we expand the operator $\hat{\Omega}$ as an (even) power series in $\nu$, with coefficients
 $c_{2n}(t)$ so that the operator may be written (in explicitly Hermitian form)
 \beq \hat{\Omega} \ = \ \sum_{n=0}^{\infty} \sqrt{c_{2n}(t)}
 \left(-i \frac{\partial}{\partial t} \right)^{2n}  \sqrt{c_{2n}(t)}  \label{eqb} \eeq
and take the asymptotic form for the Airy function
\footnote{ The Airy functions $A_i$ and $B_i$ are linear superpositions of the functions denoted here
by $A_{\pm}$.}
 \beq A_\pm(z) \ \approx  
\frac{1}{2\sqrt{\pi}(-z)^{1/4}} 
 \exp\left(\pm i \int^t \nu_\omega(t^\prime) dt^\prime \right)   \label{c}. \eeq
Using (\ref{eqb}) and (\ref{eqc}), inserting into (\ref{eqd}), and keeping only the
terms which are non-negligible in the semi-classical approximation (as described above)
we obtain
\begin{eqnarray} & & 
\left(\omega - \sum_{n=0}^\infty c_{2n}(t) \left[
 \nu_\omega^{2n} - 2n \, i \nu_\omega^{(2n-1)} \frac{d}{dt}
 \ln\left( \frac{\sqrt{c_{2n}(t)}N_\omega(t)}{(-z(t))^{1/4}} \right) 
   - i \frac{2n(2n-1)}{2}  \nu_\omega^{(2n-2)} \dot{\nu}_\omega(t) \right] \right) \nonumber \\ & &
  \ \ \  \cdot \, N_\omega(t) A_\pm(z(t)) \ = \  0 . \label{deq} \end{eqnarray}

Performing the resummation, the first two terms of eq.(\ref{deq}) cancel by virtue of
 (\ref{classical}), and the remaining terms lead  to
\beq \frac{\partial}{\partial t}  \Omega^\prime(t,\nu_\omega)
  + \Omega^{\prime\prime}(t,\nu_\omega) \dot{\nu}_\omega
  + 2 \Omega^\prime(t,\nu_\omega) \frac{d}{dt}\ln\left( \frac{N_\omega(t)}{(-z)^{1/4}}\right)
 \ = \ 0 \eeq
 or
\beq \frac{d}{dt} \ln \left(\frac{ \sqrt{\Omega^\prime} N_\omega}{(-z)^{1/4} } \right) \ = \ 0 \eeq
with solution
 \beq N_\omega(t) \ = \ \frac{(-z(t))^{1/4}}{\sqrt{\frac{1}{2}
\left|\Omega^\prime(t,\nu_\omega)\right|}}, \label{eqk} \eeq
where the  overall constant has been chosen to match the normalization constant for $t \, \sim \, t_c$
given by eq.(\ref{eqj}). This establishes the relation (\ref{eqi})
   with $N_\omega$ given by eq.(\ref{eqk}),  
 both  far away  from the critical value, $t_c$, and in the region near $t_c$.

\subsection{Green function of the BFKL operator}
We can now derive  the Green function of the BFKL operator starting from the Green function of the Airy operator,  $G_\omega(z,z^\prime)$, 
 \beq \left(z-\frac{\partial^2}{\partial z^2}\right) G_\omega(z,z^\prime) \ = \ \delta(z-z^\prime).  \label{eqp} \eeq

 Owing to  asymptotic freedom,  the BFKL scattering amplitude should tend  to zero when  $z(t) \,  \stackrel{t \to \infty} \to \, \infty $, which leads to 
the ultraviolet boundary condition
 \beq G_\omega(z,z^\prime) \stackrel{z,z^\prime \to \infty}{\rightarrow} 0 \label{eqq} \eeq
From the Wronskian of the two independent Airy functions
\beq A_i(z) \frac{d}{dz} B_i(z) - B_i(z) \frac{d}{dz} A_i(z) \ = \ \frac{1}{\pi}, \eeq
we see that a solution to eq.(\ref{eqp}) with the required ultraviolet behaviour is given by
 \beq G_\omega(z,z^\prime) \ = \ \pi\left( B_i(z) A_i(z^\prime) \theta(z^\prime-z)
 + A_i(z) B_i(z^\prime) \theta(z-z^\prime) \right) \label{eq9} \eeq
However, eq.(\ref{eq9}) is {\it not} a unique solution to eq.(\ref{eqp})
since we may add to it
any solution of the homogeneous equation  with the required  ultraviolet boundary condition,
i.e. a term proportional to $A_(z) A_i(z^\prime)$.
The general solution is therefore
\beq G_\omega(z,z^\prime) \ = \ \pi\left( \overline{B_i}(z) A_i(z^\prime) \theta(z^\prime-z)
 + A_i(z) \overline{B_i}(z^\prime) \theta(z-z^\prime) \right), \label{eq10} \eeq
with $\overline{B_i}(z)$ being the linear superposition of $A_i(z)$ and $B_i(z)$
\beq \overline{B_i}(z) \ = \ B_i(z) + c(\omega) A_i(z), \label{supp} \eeq
where $c(\omega)$ denotes a constant which only depends on $\omega$.

The function $A_i(z(t))$ has the required asymptotic behaviour as $t \, \to \, \infty$, namely
that it vanishes in that limit, whereas the function $\overline{B_i}(z(t))$ has some
 oscillatory phase for small $t$, which must match the phase of a physical wavefunction from non-perturbative
QCD, valid in the infrared region. 
Therefore, within the accuracy of the semi-classical approximation, the Green function,
 ${\cal G}_\omega(t,t^\prime)$, of the BFKL operator, with the required boundary conditions,
 can be constructed from the Green function for the Airy operator, (\ref{eq10}), allowing for the
 correcting normalization factors $N_\omega (t)$:
\beq {\cal G}_\omega(t,t^\prime) \ = \ \pi N_\omega(t) N_\omega(t^\prime) 
\left( \overline{B_i}(z(t)) A_i(z(t^\prime)) \theta(t^\prime-t)
 + A_i(z(t)) \overline{B_i}(z(t^\prime)) \theta(t-t^\prime) \right). \label{green} \eeq
From  eq.(\ref{eqi}) and using
 $N_\omega(t)$  given by eq.(\ref{eqk}), we see that 
this expression  satisfies (within the semi-classical approximation)
 the Green function equation for the BFKL operator 
\beq \left(\omega - \hat{\Omega}\left(t, -i\frac{\partial}{\partial t} \right)\right)
 {\cal G}_\omega(t,t^\prime) \ = \ \delta(t-t'),  \label{eqgre}\eeq

\subsection{Infrared boundary}
We now show how the properties of the infrared boundary determine the properties of the BFKL Green function. First we note that    
for a fixed value of $t^\prime$, the behaviour of the Green function for 
$t \, < \, t^\prime$ is controlled by the behaviour of $\overline{B_i}(z(t))$, with the oscillation phase  determined by the (non-perturbative) infrared properties of QCD.
 This  removes the ambiguity of the Green function 
given in eq.(\ref{green}) by fixing the  $\omega$-dependent constant $c(\omega$).
To see how this works we first write $c(\omega)$ in the form
 \beq c(\omega) \ = \ \cot\left(\phi(\omega)\right), \label{cot} \eeq
so that for $ t \, \ll \, t_c$, we have
 \beq \overline{B_i}(z(t)) \ \approx \ \ \frac{1}{\sqrt{\pi} z^{1/4}}   \,  
 \frac{ \sin\left(s_\omega(t)+ \frac{\pi}{4} + \phi(\omega) \right)} { \sin\left(\phi(\omega)\right)}. \label{quantcond1}\eeq
Imposing the (non-perturbative) phase condition that the argument of the sine function is 
$\eta_{np}(\omega)$  at $t=t_0$ (where  $t_0$ is small)  fixes $\phi(\omega)$  to be
 \beq \phi(\omega) \ = \ \eta_{np}(\omega , t_0)-\frac{\pi}{4}-s_\omega(t_0). \eeq 
 Note that this difference  of the  non-perturbative and perturbative phase should not depend on $t_0$.

We note, furthermore, that for the specific values of $\omega$ for which
\beq \phi(\omega) \, = \, n \pi,  \label{quantcond2} \eeq 
$c(\omega) $
 (and consequently the Green function given by eq.(\ref{green})) 
 has poles. This is to be expected
since we know that the Green function may be written in the form
 \beq {\cal G}_\omega(t,t^\prime) \ = \ 
 \sum_n \frac{f_n(t) f_n^*(t^\prime)}{(\omega-\omega_n)}
 \ + \ \mbox{ terms analytic in } \omega, \label{green2} \eeq
where $f_n(t)$ are the complete set of normalized eigenfunctions of the BFKL 
operator with eigenvalues $\omega_n$, subject to  the ultraviolet boundary condition 
 \beq f_n(t) \  \stackrel{t \, \to \, \infty}{\rightarrow} \ 0, \eeq
 which fixes the  phase of the oscillations at $t \le t_c(\omega_n)$.
The non-perturbative, infrared,  behaviour of QCD
 determines the phase of the oscillations at the infrared boundary, $t=t_0$, which we denote   as
 $\eta_{np}(\omega_n)$.
 The  two phase conditions at $t=t_0$
and $t=t_c$ serve to determine the allowed eigenvalues, $\omega_n$.

 From eqs.(\ref{supp}) and (\ref{cot}) we see that
the two expressions for (\ref{green}) and (\ref{green2}) match at the poles
 if we identify
 \begin{eqnarray} f_n(t) & = & N_{\omega_n}(t) \sqrt{\frac{\pi}{\phi^\prime(\omega_n)}}
 A_i\left(z(t)\right)_{|\omega=\omega_n}   \nonumber \\ & = & 
  N_{\omega_n}(t) \sqrt{\pi \phi^\prime(\omega_n)}
 \lim_{\omega \to \omega_n} \left\{ (\omega-\omega_n) 
           \overline{B_i}\left(z(t)\right) \right\} \end{eqnarray}

The pole-part of the Green function is thus given by
   \beq {\cal G}^{\mathrm{pole}}_\omega(t,t^\prime) \ = \ 
 \sum_n  \pi N_{\omega_n}(t) N_{\omega_n}(t^\prime)
 \frac{A_i\left(z(t)\right)A_i\left(z(t^\prime)\right)}{\phi^\prime(\omega_n)(\omega-\omega_n)}, \eeq
which, apart from  slowly varying prefactors $N_{\omega_n}(t)$  and $N_{\omega_n}(t^\prime)$,
 coincides with the
 Green function used in our analyses \cite{KLRW,KLR,EKR}. 
 
In addition to the discrete spectrum $\omega_n$, with positive values of $\omega$, the Green function has a 
contribution from the continuum of states for negative values of $\omega$. 
For negative $\omega$ there is {\it no} turning point,
$t_c$,  
so the negative $\omega$ states are not quantized.
 Their continuum  gives rise to a cut of ${\cal G}_\omega$ along the negative real axis in the $\omega$-plane and could be necessary in order for the eigenfunctions of the BFKL operator to form a complete
set of functions.
The discontinuity of ${\cal G}_\omega$ for negative $\omega$ appears owing to the condensation of the poles
of $c(\omega)$ (\ref{cot}) as $\omega \, \to \, 0$.

The inverse Mellin transform of ${\cal G}_\omega$, eq.(\ref{green}), as a 
function of the rapidity  $Y$ is given by
 \beq {\cal G}(Y,t,t^\prime) \ = \ \frac{1}{2\pi i} \int_{{\cal C}} d\omega e^{\omega Y} 
 {\cal G}_\omega(t,t^\prime) , \label{contour}\eeq
where the contour ${\cal C}$ must be taken to the right of all the poles at $\omega=\omega_n$ 
which are  given by~\footnote{The sign of $n$ has been chosen here to agree with the sign convention of our previous papers\cite{KLRW,KLR,EKR}.}
 \beq
 s_{\omega_n}(t_0) \ = \ \eta_{np}(\omega_n) + \left(n-\frac{1}{4}\right) \pi, \;\;\;\;\;\; n=1,2,3 ....
 \eeq
as follows from eqs.(\ref {quantcond1} -\ref {quantcond2}).   The discrete values, $\omega_n$ are the intercepts of the individual Regge trajectories that
comprise the QCD pomeron. 
 The perturbative quantities $s_{\omega_n}(t_0)$ depend on
 the precise details of the running coupling accounting  for heavy
quarks and any possible new physics whose threshold is below $t_c(\omega_n)$,
 that determines the allowed spectrum of Regge poles.
 In addition, in the contour integral of eq.(\ref{contour}), 
there will be also contributions from the cut at negative $\omega$, corresponding to the above-mentioned
continuum of states.
 
Therefore, if the function $\eta_{np}(\omega)$ were known then the BFKL Green function would be uniquely
determined and applicable to {\it all} processes which  
are dominated
by the interaction of the QCD pomeron. In reality, however, the infrared properties of QCD are unknown
 \footnote{ In principle one might be able to extract information about these infrared phases
from lattice QCD.} and so we need to leave $\eta_{np}(\omega)$ as a free function and fit it
from the measured structure functions at low-$x$ and other available forward diffractive data.

 As well as the pole contributions for positive $\omega$
  and the cut  (for negative $\omega$),
the complete  Green function, eq.(\ref{green}), also contains the part which is analytic
in $\omega$ except for an essential singularity at $\omega=0$ (where $z(t)$ becomes infinite for
any finite value of $t$).
 This analytic part, together with the essential singularity,
 is necessary, as we shall show
below, in order to be able to match the analysis of deep-inelastic  scattering 
at low-$x$, using the discrete BFKL pomeron with a DGLAP analysis since
it
  plays an important role in the transition between very large and moderate values of~$t$. In contrast to our previous evaluations~\cite{KLRW,KLR,EKR}, in the approach presented here,  the contributions of {\it all} the poles of the Green function 
can be evaluated  using the contour integral in the complex $\omega$ plane,
 removing the necessity to account explicitly for more and more poles
 in order to improve the accuracy of the fit as was found to be necessary
 in  \cite{KLRW,KLR,EKR}.
Moreover,  the other contributions to the  Green function,
   i.e. of the cut and of the analytic part   are
automatically taken into account.

\section{Application to Deep Inelastic Scattering}
So far, we have been considering the universal Green function of the BFKL operator,
without reference to any physical process to which this Green function is to be applied
in order to determine the amplitude for that process. This means, in particular, that
the intercepts, $\omega_n$ of the discrete Regge poles which comprise the QCD pomeron,
are process-independent but are  sensitive to any physical thresholds
which may affect the running of the coupling at momenta
 $$ k \  < \ \Lambda_{QCD} \, e^{t_c(\omega_n)/2}, $$ 
(which is larger than 10 TeV for $n \, \geq \, 3$).

We now wish to apply this Green function to the case of deep-inelastic structure
functions at low-$x$, which are constructed out of the unintegrated gluon density,
 $\dot{g}(x,t)$  ($t=\ln(Q^2/\Lambda_{QCD}^2)$).
The unintegrated gluon density is given in terms of the Green function
and proton impact factor by 
\beq \dot{g}(x,t) \ = \ \frac{1}{2\pi i}\int_{\cal C}
 d\omega x^{-\omega} \int dt^\prime {\cal G}_\omega(t,t^\prime)
 \Phi_P(t^\prime). 
 \label{invmel} \eeq
The impact factor   cannot be calculated
in perturbative QCD and must be fitted to data.
It
encodes the coupling of the QCD
pomeron to the proton and it is the only quantity in the analysis which is explicitly
process-dependent.

\begin{figure}
\centerline{\epsfig{file=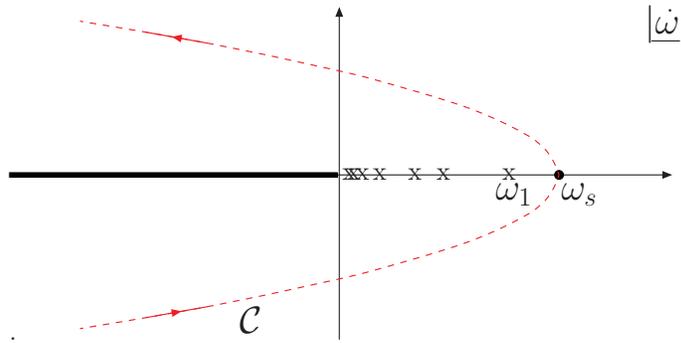, width = 9 cm}}
\caption{ The $\omega$- plane and the contour of integration of eq.(\ref{invmel})
 for the case $\omega_s(t) \, > \, \omega_1$ . The x's indicate the positions of the poles,
 which accumulate as $\omega \to 0 $ 
 \label{contour1} }
\end{figure}

 The integrand in eq.(\ref{invmel})
possesses a saddle point at $\omega=\omega_s$ in the $\omega$-plane, where 
$\omega_s$ is  given by
 \beq \frac{d}{d\omega} \ln\left({\cal A}_\omega(t)\right)_{|\omega=\omega_s} \ = \ \ln(x) \eeq
where 
 $ {\cal A}_\omega(t)$  is defined by eq.(\ref{amp}).
Provided this saddle point is also to the right of all singularities of ${\cal G}_\omega$,
i.e. $\omega_s \, > \, \omega_1$, then the contour of integration can be deformed,
as shown in Fig.\ref{contour1}, so that it passes though the saddle point in the direction
of steepest descent and the saddle-point approximation

 \beq \dot{g}(x,t) \ \approx \ \frac{1}{2\sqrt{\pi}
      \frac{d^2}{d\omega_s^2}\ln\left( {\cal A}_{\omega_s}(t)\right)}
 x^{-\omega_s(t)} {\cal A}_{\omega_s}(t) \label{sp}
 \eeq
is a good approximation to the integral over $\omega$ in the Mellin inversion eq.(\ref{invmel}).

For $t \, >\, t_c(\omega_s)$, the amplitude ${\cal A}_{\omega_s}$ has a $t$-dependence
 \beq {\cal A}_{\omega_s}(t) \ \sim \ \exp\left\{-\int^t \left|\nu_{\omega_s}(t^\prime)\right| dt^\prime  \right\}.
\eeq
For sufficiently large $t$, the classical frequency, $\nu_{\omega_s}(t)$, is approximately given by
 \beq \left| \nu_{\omega_s}(t) \right| \ \stackrel{t \to \infty}{\rightarrow} \ 
 \frac{1}{2} -\frac{\alphabar(t)}{\omega_s}
   \label{eeqqa} \eeq
and, as explained in refs. \cite{ccs,tw,abf}, in this limit the saddle-point $\omega_s$
coincides with the saddle-point obtained from inverting
 the gluon anomalous dimension, $\gamma^N_{gg}$ in the limit 
$N \, \to \, 0$. Thus we obtain a match between the BFKL analysis and the $t$-dependence
of a DGLAP analysis in the double logarithm limit, where both $t$ and $|\ln x|$ are large,
 namely (at leading order)
 \beq {\cal A}_{\omega}(t) \ \stackrel{t \to \infty}{\sim} \
     e^{-t/2} \left(t\right)^{\frac{1}{\omega \betabar}} \eeq
In this case the $t$-dependence of the unintegrated gluon density (and consequently the $t$
dependence of the structure functions) is unaffected by the discrete nature of the BFKL
spectrum.

\begin{figure}
\centerline{\epsfig{file=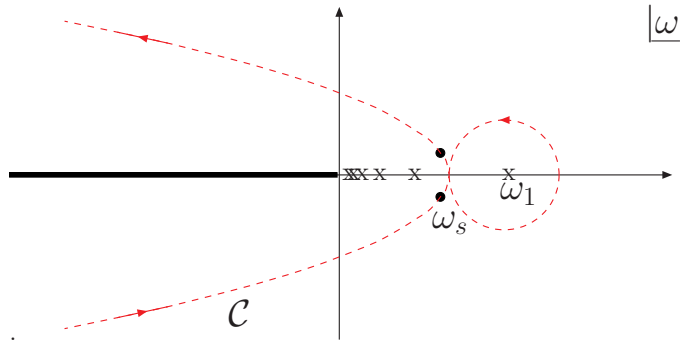, width = 9 cm}}
\caption{ The $\omega$- plane and the contour of integration of eq.(\ref{invmel})
 for the case $\Re e\{\omega_s(t)\} \, < \,\omega_1$  \label{contour2} }
\end{figure}

On the other hand, if $t$ is not sufficiently large
so that the (real  part of the)
saddle-point $\omega_s$ falls below one or more of the discrete
eigenvalues $\omega_n$, then if one attempts to deform the contour so that it passes
through the saddle-point (as shown in Fig.\ref{contour2} ) one has to surround one or
more of the discrete poles of ${\cal G}_\omega$. In this case the contribution from the saddle-point
given by eq.(\ref{sp}) has to be supplemented by the contribution from the contour surrounding
the first $j$ discrete poles for which
 $$ \omega_{j+1} \ < \omega_s \  < \  \omega_{j}, $$  
i.e. we must add the contribution

$$ \sum_{n=1}^j   x^{-\omega_n} {\cal A}_{\omega_n}(t). $$

It is these extra terms that are sensitive to the  heavy particle threshold behaviour of QCD
and which give substantial deviations to the  qualitative behaviour of the
structure functions compared with the behaviour extracted from a purely DGLAP analysis.
It is therefore important to emphasize that it is at relatively {\it low}
values of $Q^2$  and  small $x$ that we expect to see a signal of BFKL dynamics which can be
clearly distinguished from the predictions of DGLAP.

\section{Summary}
In this paper, we have analyzed the
 (Mellin transform of the) Green function for the BFKL amplitude
in the semi-classical approximation in which it can be cast into the form
 of the Green function  
 of Airy's equation after a suitable change of variables. The general solution
 contains  terms with poles for positive $\omega$ as well as an analytic part
 constructed from the  two independent solutions to Airy's equation. 
  Our expression for the Green function differs from that previously obtained
in refs.\cite{EKR,KLRW,KLR} in that in addition to the component consisting
of a set of discrete poles in the Mellin transform variable, $\omega$,
there is a component which is analytic in $\omega$. 
 This latter part turns out to be  necessary in order to
generate a match to the result of a DGLAP analysis in the DLL
limit.
 We have obtained an approximate expression for the unintegrated gluon density
by considering the saddle-point approximation to the inverse Mellin
transform. 
For sufficiently large values of transverse 
momentum the saddle-point lies to the right of all the poles and the match with the result
of a DGLAP analysis in the DLL limit follows in the same way as the case
of the continuum BFKL pomeron.  However, as the gluon transverse momentum becomes
small, the saddle-point  lies to the left of some of the discrete poles
and in such cases the unintegrated gluon density is supplemented by the
 contribution from the integral around these poles.

A complete numerical analysis, which does not rely on the saddle-point approximation,
 following the programme described in this paper is currently under 
way and will be published in a forthcoming paper.

\end{document}